\documentclass[a4paper,twoside,11pt]{article}
\usepackage[USenglish]{babel} %francais, polish, spanish, ...
\usepackage[T1]{fontenc}
\usepackage[ansinew]{inputenc}

\usepackage{lmodern} %Type1-font for non-english texts and characters

\usepackage{graphicx} %%For loading graphic files

\usepackage{caption}
\usepackage{subcaption}

\usepackage{amsmath}
\usepackage{amsthm}
\usepackage{amsfonts}

\usepackage{fancyhdr} %%Fancy headings

\usepackage[font={small,it}]{caption}

\usepackage[amssymb]{SIunits}

\newcommand{\half}{{\frac{1}{2}}}
\newcommand{\abs}[1]{{\left|{#1}\right|}}
\newcommand{\ds}{{\mathrm{d}s}}

\newcommand{\dx}{{\mathrm{d}x}}
\newcommand{\dphi}{{\mathrm{d}\phi}}
\newcommand{\dddt}[1]{\frac{{\mathrm{d}}^2#1}{\mathrm{d}t^2}}

\newcommand{\atan}[2]{\mathrm{atan2}\left({#1},{#2}\right)}

\newcommand{\norm}[1]{{\left\|{#1}\right\|}}
\newcommand{\dt}{{\mathrm{d}t}}

\newcommand{\ms}{{\meter\per\second\ }}

\begin{document}

\pagestyle{empty} %No headings for the first pages.

%% Title Page %%%%%%%%%%%%%%%%%%%%%%%%%%%%%%%%%%%%%%%%%%%%%%%
%% ==> Write your text here or include other files.

%% The simple version:
%\title{State Estimation for Airborne Wind Energy Converters - A Comparison of Cost-Efficient Sensor Setups for Control and Wind Assessment}
%\title{Cost-Efficient Sensor Setups for Control and Wind Assessment for Airborne Wind Energy Converters}
%\title{Cost-Efficient Sensor Setups for State and Wind Estimation for Airborne Wind Energy Converters}
\title{Sensor Setups for State and Wind Estimation for Airborne Wind Energy Converters}
%\author{Maximilian Ranneberg\thanks{Maximilian Ranneberg is with the Faculty of Fluid Mechanics at the TU Berlin (\emph{m.ranneberg@mailbox.tu-berlin.de}) and since 2011 engaged in Simulation and Control at EnerK\'ite}}
\author{Maximilian Ranneberg\thanks{\emph{m.ranneberg@mailbox.tu-berlin.de}}}
\date{} %%If commented, the current date is used.

%% The nice version:
%\input{titlepage} %%You need a file 'titlepage.tex' for this.
%% ==> TeXnicCenter supplies a possible titlepage file
%% ==> with its templates (File | New from Template...).

%\begin{multicols}{2}
\maketitle
%\end{multicols}

%% Inhaltsverzeichnis %%%%%%%%%%%%%%%%%%%%%%%%%%%%%%%%%%%%%%%
%\tableofcontents %Table of contents
%\cleardoublepage %The first chapter should start on an odd page.
\pagestyle{plain} %Now display headings: headings / fancy / ...
\begin{abstract}
An unscented Kalman filter with joint state and parameter estimation is proposed for aerodynamics, states and wind conditions for airborne wind energy converters.
The proposed estimator relies on different measurement setups.
Due to the strict economic constraints of wind energy converters, the sensor setups are chosen with minimal cost and reliability issues in mind.
Simulation data with a high fidelity system model and experimental tests using flight data, together with wind measurements obtained using a lidar system for altitude wind measurements, are used for validation. 
The data was obtained during test flights of the EnerK\'{i}te EK30, an airborne wind energy converter currently in research operation in Brandenburg, Germany.
Feasible accuracies were achieved even with the simplest of setups and illustrate the gain achievable by airborne sensors.
Additionally, the results encourage further research into use of the obtained wind estimates for site assessment.
\end{abstract}

\section{Introduction}
Airborne wind energy converters are a promising concept for efficient wind energy conversion.
%With the ability to reach high altitudes without costly and ecologically questionable towers, the interest in these systems is rising.
With the ability to reach high altitudes without towers, the interest in these systems is rising.
%In fact, in recent years research into this technology has increased tremendously.
% Zitieren/Labern
EnerK\'{i}te\footnote{http://www.enerkite.com} is developing airborne wind energy converters and is currently operating the research platform EK30\cite{ekBook}, for which the presented method was developed.
The EK30 is a Yo-Yo airborne wind energy systems, an idea described for example in \cite{Loyd}.
These systems fly a crosswind motion to generate high traction and reel out a tether connected to a generator.
At some point, the tether needs to be reeled in again and the kite is brought into a state that results in a positive net power generation over the full cycle.
%The results presented here were developed for and with this system.
The company SkySails has been using kites for ship propulsion and is currently developing a prototype for energy generation\footnote{http://www.skysails.info/english/power/power-system}.
Makani power\footnote{http://www.makanipower.com} are developing systems with airborne electricity generators.
At OPTEC\footnote{http://www.kuleuven.be/optec/} numerous optimization papers with applications to kite power have been published, for example \cite{Leuven1},\cite{houska1},\cite{Leuven2},\cite{Leuven3}.
A research group\footnote{http://www.kitepower.eu/} at the TU Delft have been researching airborne wind energy since 1999\cite{TUL1999} and have been focusing in the aero-elastic modeling of kites\cite{TULaer1},\cite{TULaer2} and control\cite{TULCtrl}.
They also built a prototype system for control applications.

While the material requirements are drastically lower than those of conventional wind turbines, the operational management and control requirements are significantly higher.
The airborne system needs to be actively controlled to track a desired trajectory.
In case of failures or weather conditions too harsh for the system to withstand, a safe landing needs to be accomplished.

Reliable and accurate state estimates are necessary.
For operational management and optimized control, wind condition estimates result in significant advantages.
Direct measurement of the wind speeds at kite height using a different measurement system, for example a lighter-than-air based anemometer or a lidar system, is not economically viable.
High accuracy measurements on the kite are feasible, but lead to increased cost of operation and, consequently, in increased cost of electricity.

In this article a state estimator is presented, which allows the joint estimation of wind conditions, aerodynamic parameters and system states. An unscented Kalman filter algorithm is employed, with different sensor setups including airborne sensors and ground based measurements.
\begin{itemize}
	\item Ground-based measurements (low cost),
	\item Additional acceleration and airspeed pressure measurements (medium cost, additional need of transmission with reliability issues),
	\item Additional GPS velocity and position measurements (additional high cost and additional reliability issues).
\end{itemize}
The model is presented and the estimator tested in a more detailed simulation environment.
Effects of modeling error on the state and parameter estimation are discussed for the different setups.
Using real flight data together with lidar measurements of the wind conditions in altitudes up to 200 m validates the reliability and accuracy in real-world conditions.
%While the estimator is used to identify the state of the kite, the focus of the experimental results is the aspect of wind estimation, to compare the estimator results with wind measurements.
The focus of the experimental results is the aspect of wind estimation, to compare the estimates with wind measurements.

Wind speed estimation is an important aspect for aviation and wind energy in general.
If ground speed vectors and airspeed vectors are available, the wind vector can be calculated by subtracting one vector from the other.
The airspeed measurements are actually pressure sensors, which leads to significant errors for low airspeed, even though they are accurate for speed estimation of high-speed vehicles.
%Let us assume a flight path perpendicular to the wind speed.
For a kite flying with 30 \ms perpendicular to the wind, the total airspeed difference between a wind speed 0 \ms and 10 \ms is just 1.6 \ms.
Considering the noise of common airspeed sensors for low airspeed, wind speed estimation from direct airspeed measurements is prone to noise and results in low accuracies.

Model based estimates can alleviate the problem of accuracy in pressure sensors in part.
A kite is unable to fly at 0 \ms wind speed, and in general will fly at maximal crosswind speeds of about glide ratio times wind speed without reeling out.
Thus, a model based approach is significantly more sensitive to wind speed and, consequently, allows for more accurate wind speed estimates.
Aerodynamic models are needed for a model based approach. 
In \cite{unmannedVehicle} the wind conditions are being estimated using an aerodynamic model in combination with the differential approach described above.
The wind fields for efficient soaring are calculated, with very good results in case of high accuracy airspeed vector measurements.
In \cite{WindSpeedTurbine} the aerodynamic model of a variable-speed wind turbine is used to maximize the power output and estimate the wind speed without relying on additional anemometers.
For kite control, \cite{ukfOckels} presents a filter model, a control strategy and preliminary numerical results.
A simplified aerodynamic model is used for simulation and estimation, but a more detailed tether model is used for simulation.
In \cite{EstNorway} an unscented Kalman filter using ground-based measurements is employed to estimate state and wind conditions. 
The simulation and estimation model are identical and the focus lies on the comparison of the unscented Kalman filter with the extended Kalman filter and on the influence of a wind shear model on accuracy.

In this article, joint estimation of aerodynamics, states and wind conditions is applied.
%We focus on the simulation of Yo-Yo airborne wind energy systems, an idea described for example in \cite{Loyd}.
%The system is simulated using a detailed three tether system with more realistic and complex models for tethers, actuators and aerodynamics.
Simulation data is obtained using using a detailed three tether system of the EK30 with more realistic and complex models for tethers, actuators and aerodynamics.
Accuracies during different wind conditions are presented, and experimental data is used to validate the wind estimates.
%The experiments were conducted using the research platform EnerK\'{i}te EK30 currently in operation in Germany \cite{ekBook}.

\section{Estimator Model and Filter}
In case of the minimal sensor measurement setup, only the angles and forces of the tether lines are available and the states of a full rigid-body model are highly under-determined.
Thus, a point-mass model augmented with aerodynamic coefficients for state estimation is employed.
The model is similar to the estimator and (optimal-) control models used in \cite{ukfOckels},\cite{houska1}.
%Instead of solving a differential-algebraic model, we will reduce the problem to an ODE.

The most important aspect on kite modeling is the aerodynamic forces, which are ultimately quadratic in airspeed.
An unscented Kalman filter \cite{ukf} handles nonlinearities in state update equations better than the extended Kalman filter.
The estimator is implemented in square-root formulation \cite{sqUkf}.
Since the focus of this paper is the aspect of different sensor setups and experimental results, no details are given on the implementation details and theoretical intricacies of the unscented Kalman filter.

\subsection{Aerodynamic Forces}
The aerodynamic forces are defined as follows.
Let $A$ be the aerodynamic area of the kite and $\rho$ the density of the air.
Airspeed is given by $v_a=w-v$, with the wind vector $w$ in Cartesian coordinates and the kite velocity $v$ in the same system.
The drag force is always parallel to this vector and given by
\begin{equation}
F_D = \half\rho A c_D \abs{v_a} v_a.
\end{equation}
The lift force is perpendicular to this drag force.
The notion of rolling, like an airplane, is used to integrate the effect of steering inputs.
Assuming negligible sideslip, the roll axis of the kite aligns with the airspeed. 
An initial lift vector points perpendicular to the airspeed but parallel to the tether. 
This can be interpreted as a kite without line length differences with respect to the main tether line.
The control input then results in a lift vector $Z$ by a rotation of this vector around the airspeed axis.
%Using the rodriguez formula, the resulting force direction $Z$ can be efficiently calculated.
The force is then given by
\begin{equation}
F_L = c_L \half\rho A c_L \abs{v_a}^2 Z(c_u u),
\end{equation}
where the coefficient $c_u$ describes the linear relation between the control input $u$ and the roll angle.

The tether drag force acting on the kite is integrated over every point $s$ between the ground station at length 0 and the kite at length $L$, given by 
\begin{equation}
F_{Ds} = \half\rho d c_{Ds} \int_0^L{\abs{w-\frac{s}{L}v} ({w-\frac{s}{L}v}) \ds},
\end{equation}
where $d$ is the effective diameter and $c_{Ds}$ is the drag coefficient of the lines.

%\slshape
%Note that this aerodynamic control model is comparable to different models of aerodynamic control.
%Another suitable model for ramair kites is as follows: There are two surfaces on an arc-shaped softkite, the left half and the right half of the kite.
%Both result in an aerodynamic force vector perpendicular to an effective angle on the arc.
%Without controls, the forces perpendicular to the tether direction cancel out.
%If one line is pulled, the surface on this side gains area, and thus the effective angle of both areas shift towards the pulled side.
%This can also be described with a rotation of the initial arc around the roll axis.
%\normalfont

\subsection{Lagrangian and Dynamic Equations}
The dynamic equations are derived using the system Lagrangian. The following kinetic energies are considered: the kite, with the kinetic mass of the kite $m_k$ and the encapsulated air, the tether with its density $\mu$ and the winch with rotational velocity $\omega$ and moment of inertia $J$.

\begin{align}
E_{kin} &= \half ({m_k + \rho V}) v^2 + \half \int_0^L \mu \frac{s^2}{L^2} v^2 \ds + \frac{1}{2} J \omega^2 \\
				&= \half ({m_k + \rho V + \frac{1}{3} \mu L}) v^2 + \frac{1}{2} J \omega^2
\end{align}
The work and potential energy is given by the gravitational work of the kite and the tether, the forces acting on the kite and the work done on the tether by the motor with torque $T$.
%Note that this torque is the sum of the motor torque and the friction torque.

\begin{align}
E_{pot} &= m_k z + \int_0^L \mu \frac{s}{L} z \ds \\
				&= (m_k+\half \mu L) z \\
dW      &= (F_L+F_D+F_{Ds})\dx + T\dphi
\end{align}
Since the movement of the kite is constrained by the tether, the constraint equation is given by $g(x,L) = \half (\abs{x}^2 - L^2)$ with the gradients $\nabla_x g = x^T$ and $\nabla_L = L$.
The equations are given such that the Lagrange multiplier $\lambda$ results in a tether traction force in Newton, using the radius of the drum $R$ and the normalized position $\hat r$. The twice differentiated constraint equation is given by $x^T\ddot{x} = L \ddot{L}$

\begin{align}
({m_k + \rho V + \frac{1}{3} \mu L}) \dddt{x} + \lambda \hat r &= F_L+F_D+F_{Ds}\\
\frac{J}{R} \dddt{L} - R \lambda &= T
\end{align}
Since $x^T\ddot{x} = L \ddot{L}$, with $m = {m_k + \rho V + \frac{1}{3} \mu L}$ and the identity matrix $\mathbb{I}_3$ this results in 

\begin{align}
\begin{pmatrix}
m \mathbb{I}_3 & \hat r \\
\frac{J}{R} {\hat r}^T & R
\end{pmatrix}
\begin{pmatrix}\dddt{x} \\ \lambda \end{pmatrix} &= \begin{pmatrix} F_L+F_D+F_{Ds} \\ T \end{pmatrix} \label{eq:ddx}\\
\end{align}
The structure of the mass matrix on the left hand side permits a convenient inversion. With $\tau = J+mR^2$, $\bar J = \frac{J}{m}$ and $\hat r = (x,y,z)$ the inverse is given by

\begin{align}
\frac{
\begin{pmatrix}
		R^2 \mathbb{I}_3
		+
		\bar J \begin{bmatrix}
		y^2+z^2 & -xy & -xz\\
		-xy & x^2+z^2 & -yz\\
		-xz & -yz & x^2+y^2\\
		\end{bmatrix}
		&
		R \begin{bmatrix}x\\y\\z\end{bmatrix}\\
		J \begin{bmatrix}x&y&z\end{bmatrix}
		& -Rm
\end{pmatrix}}
{\tau}
\end{align}
Note that for $J\to\infty$ the constraint force is given by $F_S = \hat r \cdot \sum F$, which is the force resulting from a fixed rod.
For $J\to 0$ the tension is given by $F_S = -T/R$, that is the force depends completely on the applied torque.

\subsection{Implementation}
Joint estimation of the parameters is employed, using the following states with additional parameters $w,c_L,c_D,c_u$. The wind $w$ is a two-dimensional vector, neglecting vertical wind speeds.
\begin{equation}
q = (x,\dot{x},\ddot{x},w,c_L,c_D,c_u)
\end{equation}
The state update equations are given by \eqref{eq:ddx}. The differential equations for the parameters are $\dot{w}=\dot{c_i}=0$. The measurement equations are dependent on the sensors available.

Tether angles are measured on the ground using angle measurement sensors and given by $\phi=\atan{y}{x}$, $\theta=\arcsin{\frac{z}{r}}$. 
The measured forces on each of the three lines are being added to be comparable to the model tether force. 
%Accelerations need to be transformed into cartesian coordinates and corrected orientation changes to be comparable to the model accelerations. 
The velocity and position is obtained using the GPS information. 
An airspeed sensor is able to obtain absolute airspeed pressure estimates.

The sigma points (see\cite{ukf}) are chosen using the parameters $\alpha=0.01,\beta=2$ and $\kappa=0$.

%\begin{table*}
	%\centering
		%\begin{tabular}{l r|l r}
		%\hline
		%Model State & Innovation Root & Measurement & Standard Deviation\\
		%\hline
			%Position & 0 m & Length & 0.5 m\\
			%Velocity & 0.1 m/s & Angles & 0.5\textdegree\\
			%Acceleration & 50 $m/s^2$ & Force & 100N \\
			%$c_L$ & 5e-4 & Acceleration & 1 $m/s^2$ \\
			%$c_D$ & 1e-5 & Airspeed Pressure ($v_a^2$) & 50 $m^2/s^2$\\
			%$c_u$ & 5e-5 & GPS Position& 3 m\\
			%Wind  & 2e-4 & GPS Velocity& 1 m/s\\
		%\hline
		%Model Parameter & & Sigma Points (see\cite{ukf}) &\\
		%\hline
			%Drum inertia & 27 $kgm^2$ & $\alpha$ & 1e-2\\
			%Aerodynamic area & 12.8 $m^2$ & $\beta$ & 2\\
			%Tether mass per length & 1.3 kg/100m & $\kappa$ & 0\\
			%Kite mass & 6 kg \\
			%Tether drag coefficient & 1.2\\
			%Tether diameter & 9 mm \\
			%Tether mass & 1.4 kg/100m
		%\end{tabular}
		%
	%\caption{Filter parameters used in the simulation runs. Note that the system is operated at 100Hz.}
	%\label{tab:PropoertiesForFilter}
%\end{table*}

%\begin{table*}
	%\centering
		%\begin{tabular}{l r}
		%\hline
		%Measurement & Standard Deviation\\
		%\hline
			%Length & 0.5 m\\
			%Angles & 0.5\textdegree\\
			%Force & 100N \\
			%Acceleration & 1 $m/s^2$ \\
			%Airspeed Pressure ($v_a^2$) & 50 $m^2/s^2$\\
			%GPS Position& 3 m\\
			%GPS Velocity& 1 m/s\\
		%\hline
		%Model Parameter & \\
		%\hline
			%Drum inertia $J$ & 27 $kgm^2$\\
			%Aerodynamic area $A$& 12.8 $m^2$ \\
			%Tether mass $\mu$ & 1.3 kg/100m \\
			%Kite mass $m_k$ & 6 kg \\
			%Tether drag coefficient $c_{Ds}$ & 1.2\\
			%Tether diameter $d$ & 9 mm
		%\end{tabular}
		%
	%\caption{
	%Filter parameters used in the simulation runs. 
	%%Note that the system is operated at 100Hz.
	%}
	%\label{tab:PropoertiesForFilter}
%\end{table*}

\begin{table*}
	\centering
		\begin{tabular}{l r|l r}
		\hline
		Measurement & Standard Deviation & Model Parameter & Value\\
		\hline
			Length & 0.5 \meter 										& Kite mass $m_k$ & 6 \kilogram \\
			Angles & 0.5\textdegree 								& Aerodynamic area $A$& 12.8 \square\meter \\
			Force & 100\newton 											& Tether mass $\mu$ & 1.3 \kilogram/100\meter  \\
			Acceleration & 1 \metrepersquaresecond 	& Drum inertia $J$ & 27 \kilogramsquaremetre \\
			Airspeed ($v_a^2$) & 50 \square\meter\per\square\second & Tether drag $c_{Ds}$ & 1.2\\
			GPS position& 3 \meter 											& Tether diameter $d$ & 9 \milli\meter\\
			GPS velocity& 1 \ms\\
		\hline
		\end{tabular}
		
	\caption{
	Filter parameters used in the simulation runs. 
	%Note that the system is operated at 100Hz.
	}
	\label{tab:PropoertiesForFilter}
\end{table*}

%\slshape
%\subsection{Using the Estimates}
%The unscented filter is advantageous, if the propagation and measurements are nonlinear.
%Then, the most likely state must not lead to the most likely value of a measurement or quantity.
%For example, the most likely wind speed value of a two-dimensional gaussion wind vector distribution is rayleigh distributed, and its most likely value will be above the wind speed of the most likely wind vector.
%Thus, we should estimate all values that are of interest with the estimated probability distribution instead of just discarding the sigma-point estimates for controller use.

%\subsection{Modelling Errors and Influence on Estimates}
%The most drastic simplifications are the tether model and the aerodynamic model.
%The tether sag and lag, due to gravitational, inertial and aerodynamic forces on the line, leads to significant errors in the position estimate, and subsequently on the velocity estimates.
%These are of vital importance to the wind speed estimates, which is still based on the airspeed at the kite and the resulting forces.
%The sag leads to assumed higher velocities, since the assumed distance is greater than the real distance. This leads to higher glide-ratios and/or higher wind speeds.
%Tether sag additionally leads to model incompatabilities. Global positiong data or estimates due to accelerometer measurements may result in significantly different, although possibly more accurate, position predictions than those obtained using angular measurements from a severly sagged line.

%\normalfont
%\newpage

\section{Simulation Model}
To validate the estimator in a simulation environment, the system model describing the research platform EK30 with the currently used wings \cite{ekBook} is employed. The model has been used for control development. Since the detailed system model is not within the scope of this article, it will only be described briefly.

\subsection{Actuator Model}
The actuator model simulates the ground station in detail. The EK30 has three coupled drums, a main drum, to which the main line is attached, and two control drums, to which the control lines are attached. Each drum and motor has coulomb and viscous friction, and external torques due to tether forces and motors are applied.

\subsection{Kite Model and Aerodynamics}
The kite is modeled using rigid-body kinematics.
The moments of inertia and the masses are defined by the used sensor unit, encapsulated air and textile material.
For orientation integration the quaternion form of the rigid-body equations is used. With the quaternion $q$ (with vector parts in the last three components), the velocity in body-fixed coordinates $v_k$, the rotational rates $\omega_k$, the earth-fixed position $x_g$, the kite mass (regarding acceleration) $m$, the inertia tensor $J$, the transformation matrix from kite-fixed coordinates to earth-fixed coordinates $T_{gk}(q)$ and the external kite-fixed forces and torques $F_k, M_k$, the equations are given by
\begin{align}
E(q) = \begin{pmatrix}
-q_1 & -q_2 & -q_3 \\
q_0 & -q_3 & q_2 \\
q_3 &  q_0 & -q_1 \\
-q_2 & q_1 & q_0
\end{pmatrix},
\end{align}
\begin{align}
\dot v_k &= \frac{F}{m} + v_k \times \omega_k  &
\dot x_g &= T_{gk}(q) v_k\\
\dot q &= \frac{1}{2} E(q)\omega_k & 
J \dot \omega_k &= M_k + J \omega_k \times \omega_k .
\end{align}
The aerodynamics are modeled using a table look-up in both airspeed angles and quadratic functions in the controls for all forces and torques in a body-fixed coordinate system.
Forces and torques due to rotational rates are modeled using linear functions in $\omega_k$.
Additional forces and torques result from the three tethers and gravitational force.
The aerodynamic functions are derived from CFD simulations and have been adjusted due to flight experience.

\subsection{Coupling and Tether Model}
The effect of control line differences is modeled more realistically than with a prescribed roll angle.
Additionally, the measured tether angles used in the filter algorithm result in errors due to sagging and lag, an effect that is captured by the more detailed model.
The three tethers are described using point masses inflexibly chained together and thus constrained in their tether-directional movement due to line acceleration. 
At each end of the tether, a spring-damper system is used to connect the actuator and the kite model.
Assuming nearly constant tension along the tether, the spring and damper constants of the material are used, but scaled according to the relation of tether length and one tether piece.
Using these assumptions, the constraint on the tether-wise direction is used to calculate the tether force on every element and consequently integrate the equations of motion for all points. This approach is similar to the method described in \cite{tetherOckels}.

\begin{figure*}[tb]
	\centering
		\includegraphics[width=\textwidth]{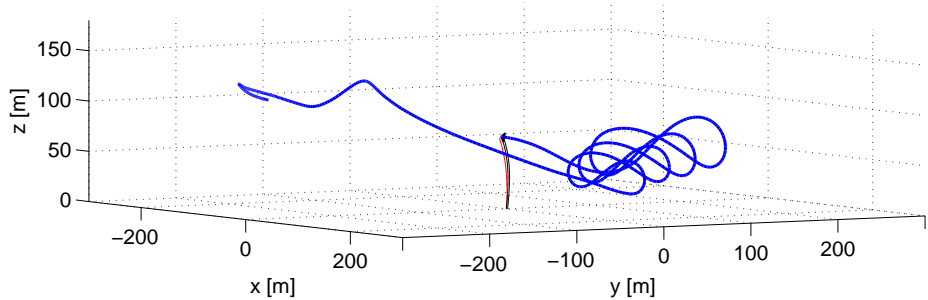}
	\caption{Typical periodic trajectory at 7 m/s wind speed.
	The trajectory is colored in blue, the main line in red, the control lines in black and the kite is shown as a blue triangle.
	%The main line does not reach the kite because the lines are attached to bridle lines, that are not shown here but included in the aerodynamic model.
	The periodic trajectory is split into two parts.
	A figure-eight path, where energy is generated by pulling out the tether, and a backtracking phase, where the kite is being flown windward to reel in with low traction force.
	The wind direction is parallel to the x-axis, with turbulence in horizontal directions.
	%With higher wind speeds, the backtracking phase takes more time while the reel-out time decreases.
	}
	\label{fig:Trajectory}
\end{figure*}

\section{Simulation Results}
The results of this sections are obtained using the model described in the previous section.
Model parameters for estimation are shown in Tab. \ref{tab:PropoertiesForFilter}.
The kite is controlled to fly figure-eight trajectories during reel-out, and to fly windward to reel-in.
A trajectory of a flight path is shown in Fig. \ref{fig:Trajectory}.
During all simulations a turbulence of 5\% was added to the wind.
The turbulence has only minor effects on the estimation quality within bounds of sensible operation but affects control accuracy, which is not discussed here.

%\begin{itemize}
	%\item What is the influence of different sensor setups on state and wind etimation?
	%\item How accurate are the results obtained with the simplified estimation model?
%\end{itemize}

A simplified point-mass model is used to explain the results of a complex aerodynamic and mechanical system.
Thus, the identified aerodynamic properties cannot be compared directly to simulation parameters.
The states of position and velocity are necessary for control applications, and wind vector and airspeed are useful for advanced control strategies.
%We will use the error quantities given below, where the filter estimates are given by $\hat\cdot$ 
%\begin{align*}
%e_{vel} &= \norm{\hat v-v}\\
%e_{pos} &= \norm{\hat x-x}\\
%e_w     &= \norm{\hat w-w_{xy}}.
%\end{align*}

%%
%\begin{table}
	%\centering
		%\begin{tabular}{l|r}
		%\hline
			%Drum inertia & 27 $kgm^2$ \\
			%Aerodynamic area & 12.5 $m^2$ \\
			%Tether mass per length & 1.3 kg/100m\\
			%Kite mass & 9.2 kg \\
			%Tether drag coefficient & 1.2\\
		%\hline
		%\end{tabular}
	%\caption{Properties for estimation model}
	%\label{tab:ProportiesForEstimationModel}
%\end{table}

\subsection{Wind Step Response}
\begin{figure*}
\centering
\begin{subfigure}[b]{\textwidth}
		\includegraphics[width=\textwidth]{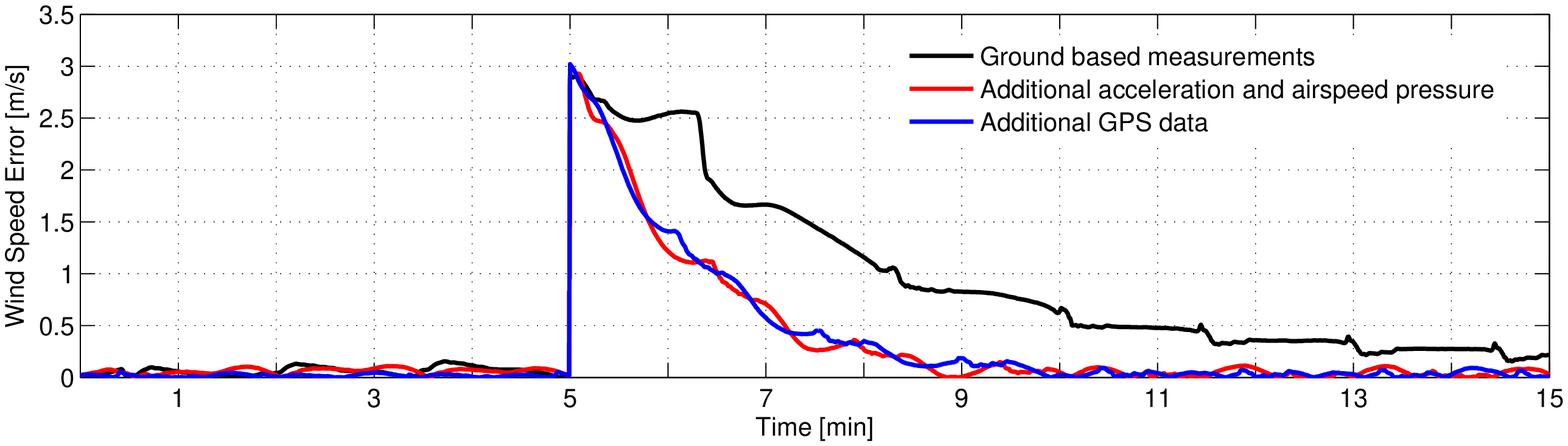}
		\caption{Wind speed error.}
	\label{fig:StepSpeed}
\end{subfigure}

\begin{subfigure}[b]{\textwidth}
		\includegraphics[width=\textwidth]{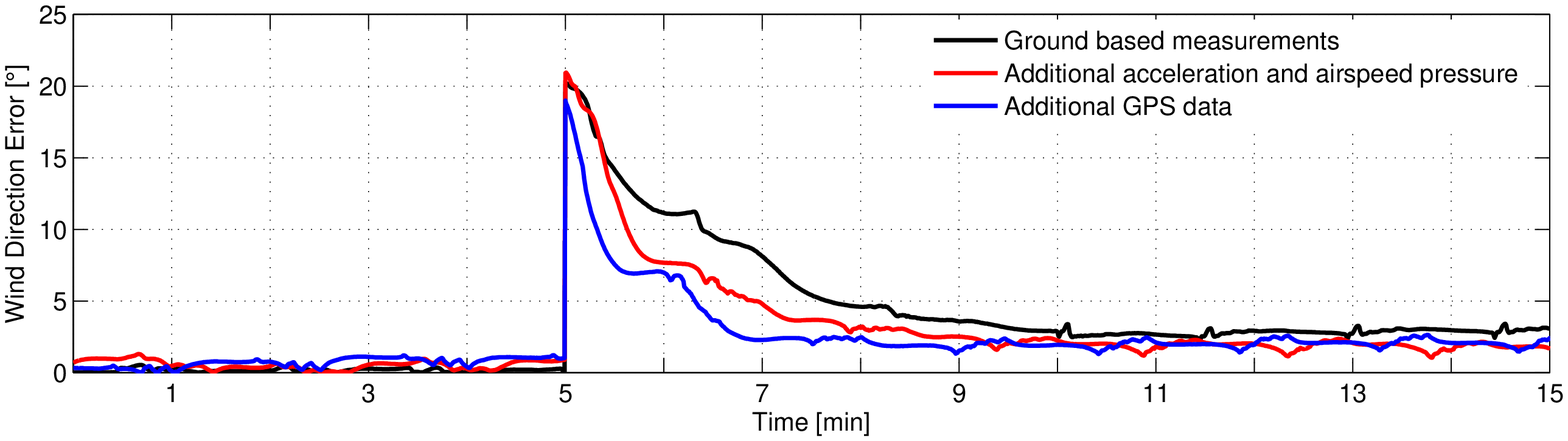}
		\caption{Wind direction error.}
	\label{fig:StepDirection}
\end{subfigure}

\begin{subfigure}[b]{\textwidth}
		\includegraphics[width=\textwidth]{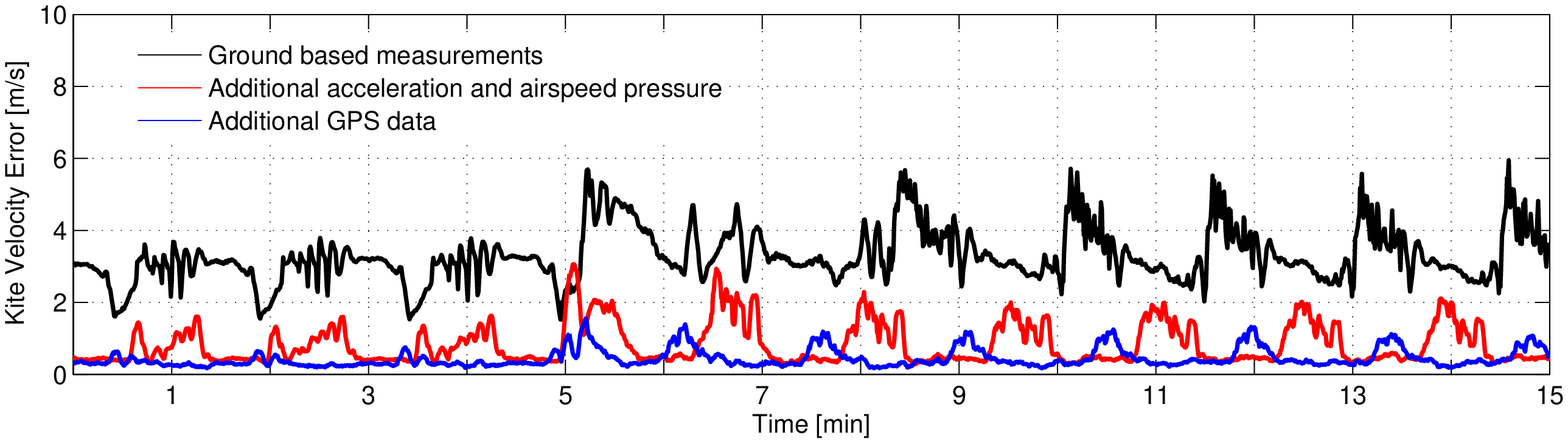}
		\caption{Kite velocity error $\norm{\hat v-v}$.}
	\label{fig:StepVelocity}
\end{subfigure}

\caption{Simulation results during a step change in wind direction and speed.}

\end{figure*}

In this subsection, a step change on the wind conditions is applied.
After 5 minutes, the direction changes about 20\textdegree and the speed increases from 7 \ms to 10 m/s.
The errors in wind speed and direction are shown in Fig. \ref{fig:StepSpeed} and Fig. \ref{fig:StepDirection}, respectively.
Wind speed errors below 0.5 \ms are achieved after 2 minutes for airborne sensors and 5 minutes for ground based estimation.
The new wind direction is found faster. For errors below 5\textdegree ground based estimates take 3 minutes and additional acceleration and airspeed pressure setups take 2 minutes. 
%Wind direction estimates are driven by symmetries, higher wind speeds means only very few symmetric figure eights and long, asymmetric backtracking phases.
%This results in more bias for higher wind speeds.
Additional GPS data results in accurate direction estimates in 1.5 minutes.
In contrast to wind speed, a gain in wind direction accuracy can be seen with GPS data.

Velocity errors are shown in Fig. \ref{fig:StepVelocity}.
During the controller adaptation time, the error graphs differ from the periodically repeated curves past 9 minutes and before the step change.
With additional accelerometer and pressure sensors, the velocity error is reduced significantly. 
Since GPS allows direct measurement of the velocity, the error can be reduced even further.

\subsection{Limit Accuracies}
\begin{figure*}

	\centering

	\begin{subfigure}[b]{\textwidth}
	\includegraphics[width=\textwidth]{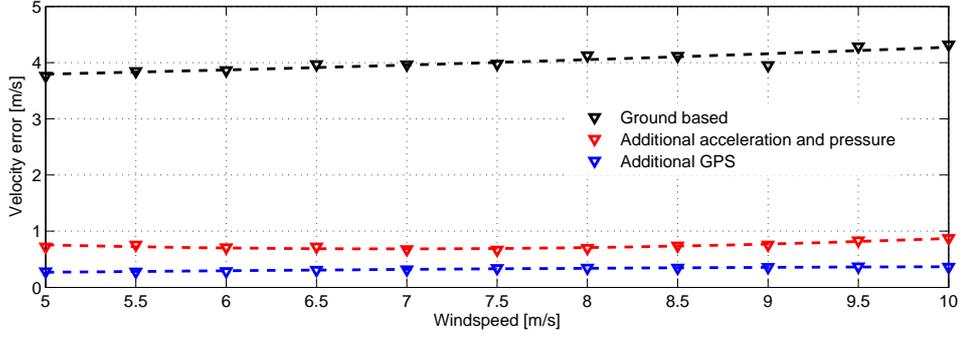}
	\caption{
	Velocity estimation error $e_{vel}^\infty$.
	%The errors are not sensitive to wind speed besides decreasing accuracy for angular measurements alone.
	%Additional measurements lead to additional accuracy.
	%Accelerometer and pressure measurements result in errors below 1m/s.
	}
	\label{fig:ErrFinVel}
	\end{subfigure}

	\begin{subfigure}[b]{\textwidth}
	\includegraphics[width=\textwidth]{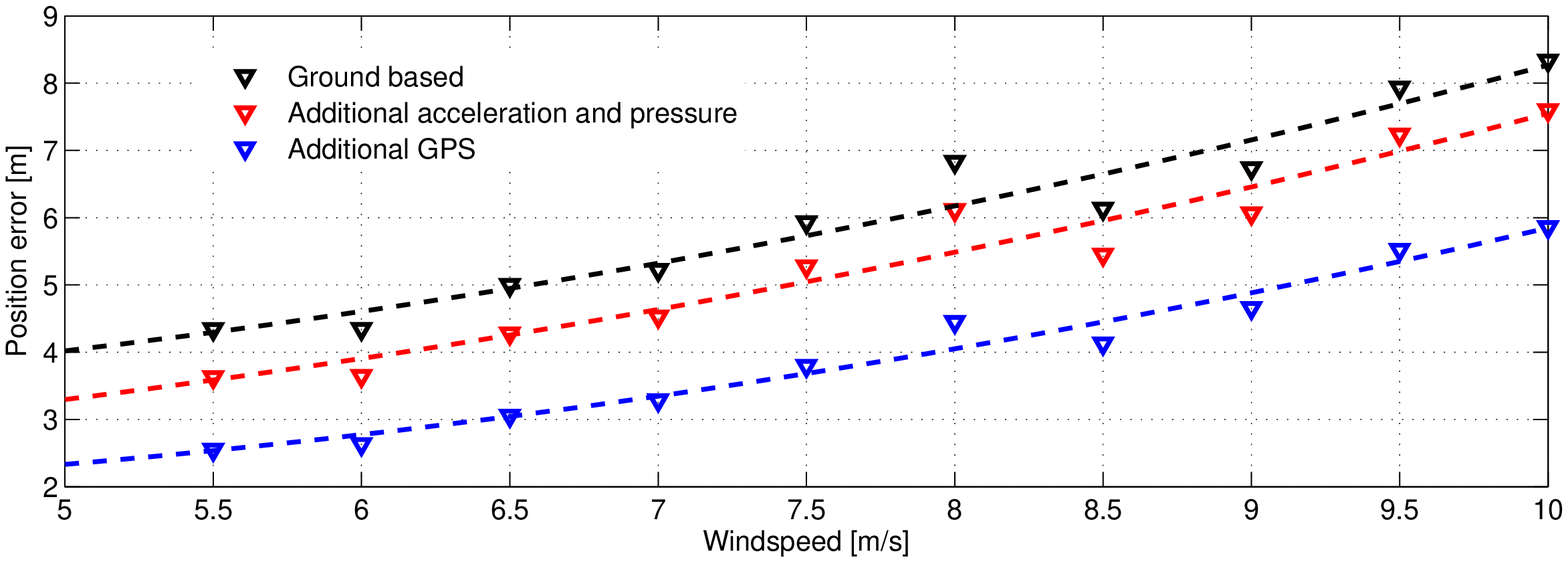}
	\caption{
	Position estimation error $e_{pos}^\infty$.
	%The higher the wind speed, the faster the movement and the shorter the periods in backtracking.
	%This leads to decreasing accuracy in position.
	%Additional information leads to additonal accuracy, especially global positioning data.
	%However, due to conflicting angular measurements and GPS position data, the error is still comparatively large.
	}
	\label{fig:ErrFinPos}
	\end{subfigure}
	
  \begin{subfigure}[b]{\textwidth}
	\includegraphics[width=\textwidth]{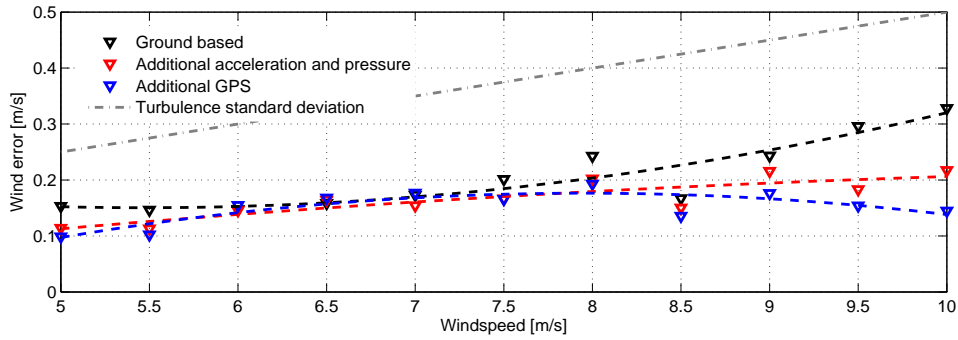}
	\caption{
	Wind estimation error $e_{w}^\infty$.
	%All measurement systems converge similar accuracies. For angular measurements alone, the accuracy decreases for higher wind speeds.
	%Additonal airborne measurements lead to increasing accuracy with higher wind speeds.
	%However, the results vary strongly depending on the final periodic trajectories and do not clearly motivate additional sensors for higher limit accuracy.
	}
	\label{fig:ErrFinWind}
	\end{subfigure}
	
	\caption{
	Limit accuracies (see equations \eqref{eq:err_final_vel}, \eqref{eq:err_final_pos} and \eqref{eq:err_final_wind}) over different wind speeds with quadratic regression fits.
	The higher the wind speed, the faster the movement and the shorter the periods in backtracking.
	%Additional accelerometer and airspeed measurements lead to significant improvements in velocity estimation, additional GPS data results in improves position accuracy.
	%Limit wind estimation errors are not sensitive to the used sensor setups at lower wind speeds.
	%Airborne measurements lead to improved accuracies at higher wind speeds.
	}
	\label{fig:ErrFin}
	
\end{figure*}

In the previous section different sensor setups during a step change in wind conditions were compared.
During constant conditions the estimation errors converge to periodic phases with the period time $T$.
The influence of different wind conditions on the accuracy can be compared using the mean error value over one period, defined in the following equations.% \eqref{eq:err_final_vel},\eqref{eq:err_final_pos} and \eqref{eq:err_final_wind}.
\begin{align}
e_{vel}^\infty &= \frac{1}{T} \int_{T} \norm{\hat v-v} \dt \label{eq:err_final_vel}\\
e_{pos}^\infty &= \frac{1}{T} \int_{T} \norm{\hat x-x} \dt \label{eq:err_final_pos}\\
e_w^\infty     &= \frac{1}{T} \int_{T} \norm{\hat w-w_{xy}} \dt. \label{eq:err_final_wind}
\end{align}
The error in velocity is shown in Fig.\ref{fig:ErrFinVel}.
Additional accelerometer and airspeed pressure measurements significantly reduce the velocity error.
Without these measurements, the velocity estimates are solely corrected due to the angular measurements that are not only lagging behind on the true movement, but misrepresent the true velocity due to the discrepancy of tether length and distance.
An effect that increases with wind speed.
There is a flat optimum at about 7.5 \ms in velocity accuracy with additional accelerometer measurements.
The negative effect of increased reel-in time and the positive effect of increased airspeed pressure result here in the most accurate estimation.
Using GPS velocity data, the error decreases further to under 1 \ms.
Due to the discrepancy between ground based measurements and GPS data, the residual could probably be decreased further by neglecting angular measurements, or increasing their uncertainty, during GPS availability.
Position errors shown in Fig.\ref{fig:ErrFinPos} are increasing with wind speed, and additional airborne sensors improve the accuracy.
%The decrease in accuracy is rather high, but so is the sagging during the reel-in phase.

The limit accuracy in wind estimation is shown in Fig.\ref{fig:ErrFinWind}.
The errors are similar across the different measurement setups.
For ground based measurement only, the errors are increasing with wind speed, but not significantly faster than the standard deviation of the turbulence rises.
Additional airborne measurements seem to rise slower than the turbulence.
However, the accuracy fluctuates more with the periodic trajectory which results at the wind speed, than with the wind speed itself.
The steady-state trajectories are not smooth functions of the wind speed.
There are only discrete numbers of figure eight paths to fly before the backtracking phase starts.
For example, at a wind speed of 7.5 \ms to 8 \ms the kite may fly 4 figure eight trajectories before the reel in.
At 9 \ms the kite may only fly 3 figure eight trajectories. 
Due to the long adaption rates of wind speed estimates, these control decisions have a significant effect on wind speed accuracy.
This stands in contrast to the other errors, where the adaption rates are short in comparison to the trajectory pattern. 
Although, the effect is hinted in the position errors in Fig.\ref{fig:ErrFinPos}, in which a step in the error values between 8 \ms and 8.5 m/s can be seen. 

%There are several factors regarding wind speed that affect estimation accuracy.
%During the reel-in phase, the kite is operated with significant tether sag and low movement speeds.
%With increasing wind-speed, the time spent in a dynamic figure-eight movement becomes smaller, while the time spent with reel-in becomes larger.
%The estimation accuracy during reel-in is worse than during reel-out, since tether sag is significantly more prominent and tether forces lower.
%At the same time, higher wind speeds hava a significantly larger effect on the airspeed in comparison the the kite speed.
%This explains the increase in accuracy with wind-speed up to a certain point, until he time spent during reel-in leads to a decrease in accuracy.

%For basic measurements, the effects of backtracking are most severe, therefore the accuracy degenerates stronger for increasing wind speeds.
%Estimating the wind speed with basic measurements relies solely on the model and, with enough time, can estimate the wind speed quite well. There is an optimal wind speed for estimation in this regard, at around 8 m/s.

%With additional airborne measurements, the wind speed is more directly correlated with the measurements and with modelling errors together result in more fluctuation and a slightly higher mean error in wind estimation during low wind speeds. For higher wind speeds, the additional airspeed measurement results in increased accuracy. 
%The error in the position and velocity estimates is generally better with additional measurements, which is also true for additional global positioning data.

%\subsection{Summary}

\begin{figure*}[tbp]
	\centering
				
		\begin{subfigure}[b]{\textwidth}
		\includegraphics[width=\textwidth]{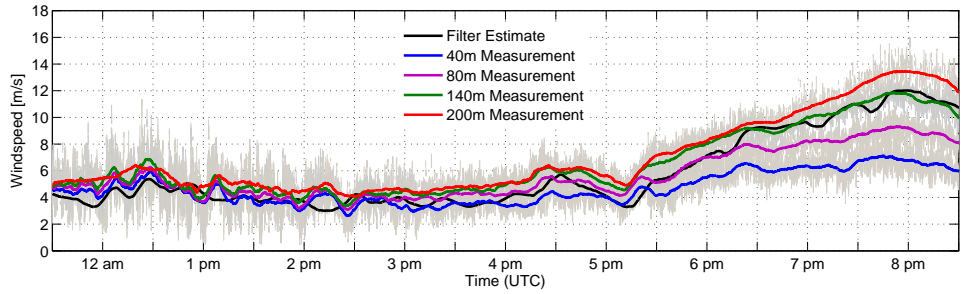}
		\caption{Ten minute mean values of the wind speeds during the flight day.}
		\label{fig:ErrExp2}
		\end{subfigure}
		
		\begin{subfigure}[b]{\textwidth}
		\includegraphics[width=\textwidth]{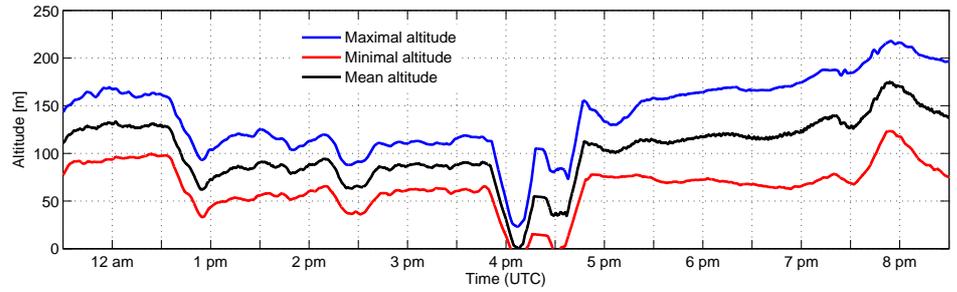}
		\caption{Operational altitudes (Min/Max/Mean over 10 minutes). The kite was landed twice past 4pm.}
		\label{fig:ErrExpH}
		\end{subfigure}
		
		\begin{subfigure}[b]{\textwidth}
		%\includegraphics[width=\textwidth]{img/iwes1.eps}
		%\caption{Wind velocities}
		%\label{fig:ErrExp1}
		\includegraphics[width=\textwidth]{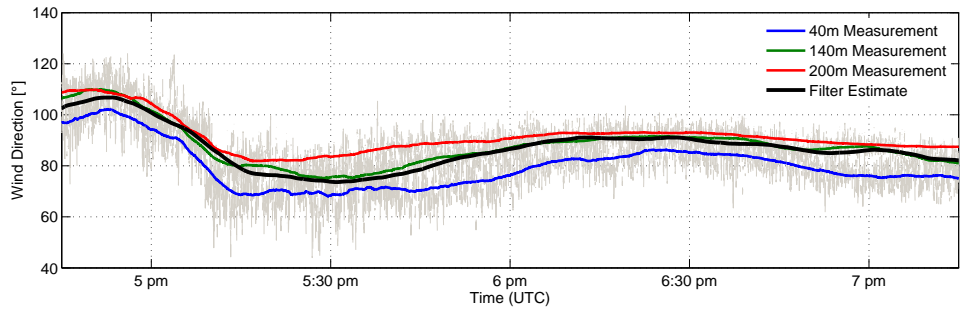}
		\caption{
		Ten minute mean values of the wind direction during the most interesting segment of the day, where the wind shear becomes apparent and the speeds rise.
		%The direction changed about 40\textdegree within 15 minutes past 5 pm and the estimate follows closely.
		%There might be a bias of a few degrees due to manual measurement of the north deviation of the filter coordinate system.
		%The absolute wind speed estimation relies on the model, but the wind direction can be inferred from symmetry considerations.
		%Thus, the direction accuracy is generally better and the adaption rates faster.
		}
		\label{fig:directionLidar}
		\end{subfigure}
		
	\caption{
	Experimental wind speed and direction.
	%on the 19th of June 2013 (Brandenburg, Germany) and corresponding operational altitude.
	%It is difficult to use an error quantity, since the wind speed depends strongly on the altitude.
	%During daytime, the wind speed is not really dependent on kite altitude, and the estimates are a bit below the range of the measurements.
	%After 3 pm (5 pm local time) the turbulence decreases and differences between altitudes become apparent.
	%The estimates closely follow the dips and peaks measured, but show additional peaks between 6 pm and 8 pm which are only hinted in the lidar values.
	%Note that the time is given in UTC, the local time ranges until 10:30 pm.
	}
	\label{fig:exp_wind}
\end{figure*}

\section{Experimental Results}
In this section results obtained during a joint flight program with the IWES Institute\footnote{http://www.iwes.fraunhofer.de} on the 19th of June 2013 are presented.
Wind speed measurements at kite height were obtained using a pulsed lidar system.
Since no additional measurements on the state of the kite or more information about the aerodynamic assumptions are available, the wind measurements are the benchmark to validate the filter estimates. 

\subsection{Setup}
No airborne measurements besides inertial accelerations were available during the course of the tests. 
The results of the estimates and measurements between noon and 22:30 are presented.
During warm and sunny summer days, the wind conditions at the site have a remarkable pattern.
With sunshine, the wind speeds during the day are low, usually under 5 m/s, with heavy turbulence especially in the z-direction. 
There are no significant differences between ground conditions and conditions at 200m altitude.
Operation during these conditions is difficult, because the tether tension needs to be very low. 
When the temperature drops during dusk, the wind speed increases over all altitudes with reduced turbulence.
Additionally, an altitude dependent profile becomes apparent, with higher wind speeds and consequently significantly higher energy densities at higher altitudes.

Pulsed lidar measurement systems allow measurements of wind speeds in variable altitudes \cite{lidar2}.
In \cite{lidar1} lidar and anemometer measurements of wind turbines at altitudes of 100m are compared and show excellent correlation.
%For airborne wind energy converters, this technology is even more important than for conventional wind turbine. 
%Knowledge of the wind speed at different heights enables more efficient energy conversion, and knowledge of the wind shear can ultimately decide on the economic sensibility of airborne wind energy at specific locations.
A Leosphere\footnote{http://www.leosphere.com/} WINDCUBEv2 was used to obtain the measurements.

\subsection{Results}
The measurements and the wind speed estimates are shown in Fig. \ref{fig:exp_wind}.
Operational altitude is shown in Fig.\ref{fig:ErrExpH}, to allow the comparison between filter estimate at operational height and lidar measurement.
During the day until 4:30pm, high turbulence and similar wind speeds across all altitudes can be seen in Fig.\ref{fig:ErrExp2}.
The wind speed estimates are similar to the measurements, but occasionally 0.5 \ms to 1 \ms lower.
When the turbulence decreases and the wind profile separates after 5:30 pm, a close following of the estimates and the measurements at operational altitudes can be seen, including the peaks and dips.
There are, however, additional dips.
The dip at 7:30 pm corresponds to a decrease in operational altitude.
This illustrates the difficulties that arise. 
Not only the trajectories greatly affect the estimation quality.
A change in aerodynamic parameters, for example because the pilot decides that a change in the difference between the control lines and the main line are in order to improve flight stability, will have an impact on the wind speed estimation.
The system takes time to adapt to the changed aerodynamics, and in between additional dips or peaks may occur.
In Fig.\ref{fig:directionLidar} the wind directions are shown during the segment of the flight, where the low wind speed turbulent conditions change to the separated conditions.
The direction estimate follows the measured values at the different altitudes, and most closely the measurement at 140m which corresponds to the mean operational altitude.

%\begin{figure}[tbp]
	%\centering
		%\includegraphics[width=0.5\textwidth]{img/iwes1Height.eps}
	%\caption{Altitudes during experimantal operation with estimates and measurements shown in Fig. \ref{fig:exp_wind}. During lower wind speed conditions during the day, the kite was flown at heights between 50m and 100/150m. During later hours the wind speed increased and the heights were increased to 100m to 150/200m from around 16:30.}
	%\label{fig:exp_wind_height}
%\end{figure}

%\slshape
%\begin{itemize}
	%\item Estimated Aerodynamics?
	%\item Estimated Wind direction?
%\end{itemize}
%\normalfont

%\subsection{Summary}
%\begin{itemize}
	%\item Good consistency between estimation and measurement of wind, but with difficulties regarding the horizontal profile. Since we operate on a range of heights and the estimation times are long.
	%\item Peaks and dips in wind speed are followed accurately.
	%\item (possible) Range of estimated aeordynamics similar to the estimates within the simulation.
%\end{itemize}

%\newpage

\section{Summary and Outlook}

%Simulation:
%\begin{itemize}
	%\item Model mismatch small enough to allow wind estimates below half a meter
	%\item Medium cost solution results in significant velocity estimation and thus advantages for control
	%\item Wind speed estimates similar, with different convergence times
	%\item Similar limit accuracies for wind speed indicate that model mismatch error and trajectory patterns are more influential than sensor availability in terms of wind speed estimates. Higher accuracy due to more sensors only in higher wind speed ranges. Conergence time dependent on sensor availability.
		%\item Even without airborne measurements, good estimates allow for efficient control
%\end{itemize}

A simple nonlinear point-mass model with aerodynamic parameter and wind velocity states for an airborne wind energy converter system was presented.
The model was used with an unscented Kalman filter to estimate the kinematic states, the aerodynamics and the current wind conditions with three different sensor setups with different cost and reliability requirements in mind.
Using a significantly more detailed simulation model including a rigid-body model of the kite, sophisticated aerodynamics, detailed actuator dynamics and a discretized elastic tether the approach was evaluated for more elaborate system models.
The limit accuracies of mean errors over steady-state periodic trajectory for constant wind speeds were used to evaluate the filter accuracy over a range of different wind speeds.
The simulations show good results with respect to wind estimation even for low-cost ground-based-only sensors.
Decreased reaction times can be achieved with more advanced airborne sensors.
Acceleration and airspeed pressure data result in significant improvements for velocity estimation, while GPS measurements unsurprisingly lead to higher position accuracy.

%Experiment:
%\begin{itemize}
	%\item Good match to lidar measurements in experiments
	%\item Results motivate to be confident with wind estimates and simulation results
	%\item Turbulence levels probably difficult to establish, even with airborne sensors
%\end{itemize}

%Outlook:
%\begin{itemize}
	%\item Review using a different wing and control, with significantly shorter backtracking phase and thus increased accuracy.
	%%\item Model mismatch driving residual in wind estimation -> More detailed estimator model with full rigid-body model and aerodynamics
	%\item A true wind speed model based on a logarithmic function will probably allow a full wind profile identification in simulation and may result in reasonable profiles for real wind conditions, depending on the accuracy of the logarithmic assumption.
	%\item More detailed measurement with full airspeed vector possible
	%%\item If the lidar measurements should be possible again, operation at constant heights for estimation would allow a more thourough validation.
%\end{itemize}

The simulations show that the results depend strongly on the flown trajectory.
Long backtracking times reduce the accuracy due to strong sagging and slow movement, as well as significant aerodynamic changes due to strong sideslip.
%Additionally, though this is not discussed in the simulations section, a high sideslip angle occurs during these backtracking maneuvers which increases model mismatch.
Using different wings, a different backtracking trajectory is possible by reducing the glide-ratio significantly and directly flying directly towards the ground station.
The results will need to be reevaluated with these trajectories, at least with respect to the influence of wind speed on accuracies.

It will be possible to use full airspeed vector measurements with the EK30 in the near future.
It may allow for considerable shorter adaption times with regards to the wind speed, and thus could allow turbulence assessment.
A point mass model may not be the most efficient use of data, with a setup including inertial measurements, GPS data and airspeed vectors.

In June 2013 flights with the EnerK\'{i}te EK30 research platform were conducted, with additional lidar wind speed measurements in a range of altitudes with detailed results up to 200 \meter.
The comparison of filter estimates and lidar measurements show a close correlation.
Although it is difficult to quantify the error due to a range of altitudes of operation with significantly different local wind speeds, the estimation seems to be very close to the measured values at the mean altitude.
However, turbulence levels remain difficult to estimate from kite dynamics.
The response time of the wind estimates is between half a minute for airborne sensors to two minutes without airborne sensors, and thus short time wind variance cannot be picked up.
While airborne measurements are able to decrease this response time considerably, for reliable turbulence estimates the response time is still too slow.
The wind speed estimation is sensitive to aerodynamic changes and thus great care must be taken with the estimated wind speed values.

The good correlation with the wind speed measurements is encouraging, but the ability to fly in a range of altitudes is not efficiently used.
%If we assume a logarithmic wind profile, we only increase the wind model by a single parameter.
A common model for the local wind shear is a logarithmic profile, see for example \cite{gasch}.
However, these models are only valid up to a certain altitude.
The model-based approach may allow us to estimate the wind conditions within the bounds of the operational altitudes, as long as the wind shear model is suitable at the site.
%If the logarithmic profile is a good model for the local wind shear, this will decrease model mismatch and improve the accuracy of all state estimates.
With additional comparisons and sufficiently close correlation, these estimates may be a viable option for reliable wind site evaluation.

\section*{Acknowledgments}
This research was made possible by funds from the ILB\footnote{http://www.ilb.de/}.
The author would like to express his gratitude to the EnerK\'ite Team, especially Alexander Bormann for discussion and proofreading and Stefan Skutnik for insight into sensors.

%\newpage
%\input{appendix}

%% The List of Figures
%\clearpage
%\addcontentsline{toc}{chapter}{List of Figures}
%\listoffigures

%% The List of Tables
%\clearpage
%\addcontentsline{toc}{chapter}{List of Tables}
%\listoftables

%%%%%%%%%%%%%%%%%%%%%%%%%%%%%%%%%%%%%%%%%%%%%%%%%%%%%%%%%%%%%
%% APPENDICES
%%%%%%%%%%%%%%%%%%%%%%%%%%%%%%%%%%%%%%%%%%%%%%%%%%%%%%%%%%%%%
%\appendix
%% ==> Write your text here or include other files.

%\input{FileName} %You need a file 'FileName.tex' for this.


\begin{thebibliography}{91.}%

\bibitem{unmannedVehicle} Langelaan, Jack W., Nicholas Alley, and James Neidhoefer. "Wind field estimation for small unmanned aerial vehicles." Journal of Guidance, Control, and Dynamics 34.4 (2011): 1016-1030.

\bibitem{WindSpeedTurbine} Qiao, Wei, et al. "Wind speed estimation based sensorless output maximization control for a wind turbine driving a DFIG." Power Electronics, IEEE Transactions on 23.3 (2008): 1156-1169.

\bibitem{ukfOckels} Williams, Paul, Bas Lansdorp, and Wubbo Ockels. "Nonlinear control and estimation of a tethered kite in changing wind conditions." Journal of guidance, control, and dynamics 31.3 (2008): 793-799.

\bibitem{houska1} Houska, Boris, and Moritz Diehl. "Optimal control for power generating kites." European Control Conference. 2007.

\bibitem{tetherOckels} Williams, Paul, Bas Lansdorp, and Wubbo Ockels. "Modeling and control of a kite on a variable length flexible inelastic tether." AIAA Guidance, navigation and control conference. 2007.

\bibitem{ukf} Wan, Eric A., and Rudolph Van Der Merwe. "The unscented Kalman filter for nonlinear estimation." Adaptive Systems for Signal Processing, Communications, and Control Symposium 2000. AS-SPCC. The IEEE 2000. IEEE, 2000.

\bibitem{sqUkf} Van Der Merwe, Rudolph, and Eric A. Wan. "The square-root unscented Kalman filter for state and parameter-estimation." Acoustics, Speech, and Signal Processing, 2001. Proceedings.(ICASSP'01). 2001 IEEE International Conference on. Vol. 6. IEEE, 2001.

\bibitem{ekBook} Alexander Bormann, Maximilian Ranneberg, Peter K\"ovesdi, Christian Gebhardt and Stefan Skutnik "EnerK\'{i}te - Development of a three-line ground-actuated AWEC". In Uwe Ahrens, Moritz Diehl, Roland Schmehl, eds. "Airborne Wind Energy", Springer, 2012

\bibitem{EstNorway} Hjukse, Ola Holter. State Estimation and Kalman Filtering of Tethered Airfoils: by use of ground based meausrements. Diss. Norwegian University of Science and Technology, 2011.

\bibitem{lidar1} Smith, David A., et al. "Wind lidar evaluation at the Danish wind test site in H{\o}vs{\o}re." Wind Energy 9.1-2 (2006): 87-93.

\bibitem{lidar2} Mikkelsen, Torben. "On mean wind and turbulence profile measurements from ground-based wind lidars: limitations in time and space resolution with continuous wave and pulsed lidar systems." (2009).

\bibitem{Loyd} Loyd, Miles L. "Crosswind Kite Power (for large-scale wind power production)." Journal of Energy 4.3 (1980): 106-111.

\bibitem{Leuven1} Ilzhoefer, A., Boris Houska, and Moritz Diehl. "Nonlinear MPC of kites under varying wind conditions for a new class of large-scale wind power generators." International Journal of Robust and Nonlinear Control 17.17 (2007): 1590-1599.

\bibitem{Leuven2} Houska, Boris, and Moritz Diehl. "Robustness and stability optimization of power generating kite systems in a periodic pumping mode." Control Applications (CCA), 2010 IEEE International Conference on. IEEE, 2010.

\bibitem{Leuven3} Sternberg, Julia, Boris Houska, and Moritz Diehl. "A structure exploiting algorithm for approximate robust optimal control with application to power generating kites." American Control Conference (ACC), 2012. IEEE, 2012.

\bibitem{TUL1999} Meijaard, J. P., W. J. Ockels, and A. L. Schwab. "Modelling of the dynamic behaviour of a Laddermill, a novel concept to exploit wind energy." Proceedings of the III Symposium on Cable Dynamics. 1999.

\bibitem{TULaer1} Breukels, J., and W. J. Ockels. "Analysis of complex inflatable structures using a multi-body dynamics approach." Proceedings of the 49th AIAA/ASME/ASCE/AHS/ASC Structures, Structural Dynamics, and Materials Conference, Schaumburg, IL, USA. 2008.

\bibitem{TULaer2} Terink, E. J., et al. "Flight dynamics and stability of a tethered inflatable Kiteplane." Journal of Aircraft 48.2 (2011): 503-513.

\bibitem{TULCtrl} Baayen, Jorn H., and Wubbo J. Ockels. "Tracking control with adaption of kites." Control Theory \& Applications, IET 6.2 (2012): 182-191.

\bibitem{gasch} Gasch, Robert, and Jochen Twele, eds. "Wind power plants: fundamentals, design, construction and operation." Springer, 2012.

\end{thebibliography}
\end{document}